# QoS Routing and Performance Evaluation For Mobile Ad hoc Networks Using OLSR Protocol

Mohamed Amnai[1]; Youssef Fakhri[1,2]; Jaafar Abouchabaka[1]

[1] Laboratoire de Recherche en Informatique et Télécommunications (LARIT), Equipe Réseaux et Télécommunications Faculté des Sciences, Kenitra, Maroc
[2] LRIT Unité Associée au CNRST, Faculté des Sciences de Rabat, Maroc,
`amnai_med@hotmail.com, fakhri-youssef@univ-ibntofail.ac.ma, aboucha06-univ@yahoo.fr`

*ABSTRACT*

*Mobile Ad-Hoc network is a collection of mobile nodes in communication without using infrastructure. As the real-time applications used in today's wireless network grow, we need some schemes to provide more suitable service for them. We know that most of actual schemes do not perform well on traffic which is not strictly CBR. Therefore, in this paper we have studied the impact, respectively, of mobility models and the density of nodes on the performances (End-to-End Delay, Throughput and Packet Delivery ratio) of routing protocol (Optimized Link State Routing) OLSR by using in the first a real-time VBR (MPEG-4) and secondly the Constant Bit Rate (CBR) traffic. Finally we compare the performance on both cases. Experimentally, we considered the three mobility models as follows Random Waypoint, Random Direction and Mobgen Steady State. The experimental results illustrate that the behavior of OLSR change according to the model and the used traffics.*

## 1. INTRODUCTION

Mobile Ad-Hoc Network (MANET) is a self-configuring network of mobile nodes connected using wireless links, forming a random topology. The nodes move freely and randomly. The network's wireless topology may be unpredictable. The minimal configuration, the quick deployment and the absence of a central governing authority make ad hoc networks suitable for several positions as the multimedia teleconferences, construction site, network residence and military conflicts etc [1]-[4].

The Mobility models define nodes movement pattern in ad hoc networks. The random behaviour of these models as well as their implementations on the final ones (computer, phone…), requires some researches on the evaluation of routing protocols based on simulations.

The aim of a routing protocol is to discover the best route that links up two nodes while guarantying a QoS in communication [5]. The quick change and unpredictable of the topology of MANET network according to the random mobility of nodes, makes route research difficult to the routing protocol.

It is clear that the service quality QoS [6] in MANET is not guaranteed because of the inherent dynamic nature of a mobile ad hoc environment. In general, the performances depend on the routing mechanism and nature of mobility. In order to guarantee the QoS we should process to deepened studies of evaluation regarding to find the routing protocol and the mobility model that are more adapted to an application. The QoS call for some of the performance metrics as the throughput, the end-to-end delay and the jitter etc.  Therefore many researches were carried out on evaluation performances of the MANETs as, the performance analysis of the different routing protocols and the effect of the random mobility models on Ad Hoc networks [7]-[12].

The rest of this paper is organized as follows: In the next section, we survey related work. The problem formulation is discussed in section 3, followed by the simulation environment used in this study. The results obtained in this simulation are also discussed in section 5. In the end, section 6 completes the paper.

## 2. Related Work

In the [13] Gupta and Kumar introduced a random network model for studying throughput scaling in a fixed wireless network; the authors in the [14] have showed that at the time of movement nodes, the throughput scaling changes completely. According to [13], [14] the authors in [15] showed that the throughput and the delay are characterized by three parameters: the number of hops, the transmission range, the mobility and velocity of the node. The authors propose schemes that exploit the three features to obtain different points on the throughput-delay curve in an optimal way.

In [16] the authors showed that the delay is influenced by different network parameters: channel access probability, transmission power or radius, network load and density of nodes.

The tradeoffs delay-throughput is the object of a study for the authors of the paper [17]. The same authors developed an algorithm to achieve the optimal tradeoffs delay-throughput on certain conditions on the delay.

The effects of various mobility models on the performance of the two routing protocols (DSR-Reactive Protocol) and (DSDV-Proactive Protocol) has been studied in [1]. The four mobility models considered are: Random Waypoint, Group Mobility, Freeway and Manhattan models. The study has shown that the performances vary with the change of used mobility models.

The performance of two prominent routing protocols in MANET: OLSR and AODV are compared in the paper [18]. According to this paper, the AODV protocol will perform better in the networks with static traffic. It uses fewer resources than OLSR, because the control messages size is kept small requiring less bandwidth for maintaining the routes and the route table is kept small reducing the computational power. The AODV protocol can be used in resource critical environments. The OLSR protocol is more efficient in networks with high density and highly sporadic traffic. But the best situation is when there is a large number of hosts. OLSR requires that it continuously has some bandwidth in order to receive the topology update messages.

The authors of the paper [11] present the performance of Destination-sequenced Distance Vector (DSDV) in four different mobility models called: Random Waypoint, Reference Point Group Mobility (RPGM), Gauss Markov and Manhattan Mobility Model. In this paper, the results show that DSDV protocol with RPGM mobility model has optimized results varying network load and speed.

In the paper [19] the OLSR and Ad hoc On-demand Multipath Distance Vector (AOMDV) routing protocols have been evaluated over Levy-Walk and Gauss-Markov Mobility Models. The AOMDV routing protocol has a higher packet delivery and throughput while OLSR has less delay and routing overhead at varying node density.

In [20] the authors have used a method to evaluate performance, in terms of delay, Dynamic Source Routing (DSR) in MANET with a multi-services traffic.

It is proposed in [21] a formulation of the routing problem in multi-services MANETs as well as the implementation of an adaptation of DSR protocol.

The AODV routing protocol associated with the three mobility models (Random Waypoint, Random Direction and Mobgen Steady State) have been evaluated in the first part of [8] by using the CBR traffic. It is shown that the optimal delay is achieved by Random Way Point in weak densities of nods and by Mobgen Steady State over high density of nodes. Nevertheless, the optimal throughput is achieved by Random Way Point during the weak and big densities of nods. In the second part of paper [8] the authors analyzed the behaviour of the AODV protocol with the same previous mobility models. But this time the study is taken with a multiservice traffic (VBR MPEG-4). In this part, Mobgen Steady State outperforms Random Way Point mobility model in terms of delay over all densities of knots used. However with weak densities the optimal throughput is got by Random Way Point and when the big densities used the optimal one is represented almost by both Random Way Point and Mobgen Steady State. Generally, the AODV protocol has shown a sensitive behaviour for the type of used traffic. This change of behaviour of AODV enables to do this comparative study using a proactive routing protocol OLSR under the two types of traffic (CBR) and (VBR).

## 3. PROBLEM FORMULATION

Among the major challenges of the axes of research in the Ad Hoc networks with a density of nodes, what are the routing protocols as well as the fitting mobility models to use for a scenario of given application? What the effective parameters that gives the reliable results?

It is evident that the QoS must guarantees a certain level of performances for different applications. However, the Ad Hoc network is used in applications with different levels of QoS. The network traffic is classified into time sensitive traffic. In this category we find the applications real time traffic that requires the minimal guarantee of delay. Generally it must work without losing the data (e.g. video conferencing) [22]. Some applications in real time possess limits of the delay that must be guaranteed, but these bounds can be slightly exceeded. In this categories many application can also tolerate a small amount of packet loss [23]. The second category, it's data traffic which has no delay requirements but short average delay is desired. Data traffic requires lossless transmission [22].

From bit rate point of view, we have got two classes of traffic Constant Bit Rate (CBR) and Variable Bit Rate (VBR). In the first class some applications generate the traffic in fixed rate. As regards practicing, some applications generate a traffic CBR. In the second class most of the applications generate variable bit rate streams (VBR). This traffic is characterized by changing of the amount of information transmitted by unit time, (i.e. the bit rate). The degree of variation in bit rate is different from one application to another [25].

Some researches have focused on performances evaluation of routing protocols and the mobility models given that most of previous researches focused on traffic CBR which is not adapted to the multimedia applications of the type of traffic VBR [24].

The objective of our work is to evaluate differently the performances of OLSR routing protocol, and to study the behaviour of this protocol using the traffic CBR and VBR with different mobility models.

We have studied the impact of the nodes density on performances (End-to-End Delay, Throughput and Packet Delivery Ratio) of OLSR routing protocol. The three mobility models considered are: Random Way Point, Mobgen Steady State and Random Direction.

The VBR traffic closely matches the statistical characteristics of a real trace of video frames generated by an MPEG-4 encoder [24]. Two parameters were used to control the traffic stream. The first parameter, the initial seed, results in the variants of traffic trace. This parameter was kept constant at 0.4 [25], as the same traffic trace needed to be used in all the experiments. The second parameter, the rate factor, determined the level of scaling up (or down) of the video input while preserving the same sample path and autocorrelation function for the frame size distribution. Its value is 0.33 for 40 source, and 0.25 for 10, 20, 30 sources [23].

It is clear that the reliable of performance results is based on, the effective selection of the parameters of the simulations. In simulations of mobile ad hoc networks, the probability distribution that manages the movement of the nodes typically varies according to the time, and converges to a "steady-state" distribution. When node speeds and locations are chosen from their steady-state distributions, the parameters of performance for a given protocol, convergent towards their values to steady-state values as well. In [26], the authors show that more than 1000 seconds of simulation time may be needed to reach steady state [27]. For this reason the simulation time used in our works is 1200 seconds.

Optimized Link State Protocol (OLSR) [28] is a proactive routing protocol, so the routes are always immediately available when needed. In OLSR, each node periodically constructs and maintains the set of neighbours that can be reached in 1-hop and 2-hops. Based on this, the dedicated Multipoint Relays (MPR) algorithm minimizes the number of active relays needed to cover all 2-hops neighbours. A node forwards a packet if and only if it has been elected as MPR by the sender node. In order to construct and maintain its routing tables, OLSR periodically transmit link state information over the MPR backbone. Upon convergence, an active route is created at each node to reach any destination node in the network.

The mobility model is designed to describe the movement pattern of mobile user, and how their location, direction of movement, pause distribution, speed and acceleration change over time. The mobility models emulate a real world scenario for the way people might move in, for example, a conference setting or museum...

### 3.1. Random Way Point (RWP)

In this model, each node is assigned an initial location, a destination, and a speed. The points initial location and destination are chosen independently and uniformly on the area in which the nodes move. The speed is chosen uniformly on an interval, independently of both the initial location and destination. After reaching the destination, a new destination is selected from the uniform distribution, and a new speed is chosen uniformly on [min-speed, max-speed], independently of all previous destinations and speeds. The node stays for a specified pause time upon reaching each destination, before repeating the process [11], [25].

### 3.2. Random Direction (RD)

In the Random Direction Mobility Model each node is assigned an initial direction, speed and a finite travel time. The node then travels to the border of the simulation area in that direction. Once the simulation boundary is reached, the node pauses for a specified time, chooses another angular direction (between 0 and 180 degrees) and continues the process. The Random Direction Mobility Model was created to overcome clustering of nodes in one part of the simulation area produced by the Random Waypoint Mobility Model. In the case of the Random Waypoint Mobility Model, this clustering occurs near the center of the simulation area.

### 3.3. Mobgen Steady-State (Mbg-SS)

The implementation [19] of the RWM model with setdest for NS2, starts with a constant pause time to the initial location [29], [30]. In the other hand, the initial positions are chosen uniformly. With mobgen for NS2 [31], an other implementation of the model RWM in NS2, begins roughly by the half of the nods in movement and the second half in pause [32]. For this reason, simulations using setdest takes more time to converge to the stationary state that simulations using mobgen. When node speeds and locations are chosen from their steady-state distributions, the performance metrics for a given protocol, convergent towards their values to steady-state values as well. For this reason, at the time of the usage of setdest or mobgen, the performances network systematically can change with the time and the measures of collected performances during the convergence period cannot reflect the values in the long term [26]. The model of mobility Mobgen Steady State is an improvement of the model RWP. In this model the initial positions and the speeds of the knots are chosen from their stationary distributions. Convergence is immediate and the results of performances are reliable. The code of the model Mobgen Steady State is available to [33].

## 4. SIMULATION ENVIRONMENT

In order to achieve our aim we need to investigate how the OLSR protocol behaves when load of nodes increases with different Mobility Models (Random Waypoint, Random Direction and Mobgen Steady State). Simulations have been carried out by Network Simulator 2.34 NS-2. Multimedia traffic VBR (MPEG-4) and CBR are used. In Table 1, we provide all simulation parameters.

Table 1. Simulation parameters.

| Parameter | Value |
| --- | --- |
| Simulation Time | 1200 sec |
| Number of nodes | 10, 20, 30, 40, 50, 60, 70, 80, 90, 100. |
| Pause Time | 0 Sec |
| Environment Size | 1000 m X 1000 m |
| Traffic Type | Variable Bit Rate (VBR) MPEG-4, |
| Maximum Speeds | 10m/s |
| Mobility Models | Random Waypoint, Random Direction, Mobgen Steady-State. |

### 4.1. Performance Metrics

For the simulation results, we have selected the end-to-end delay and throughput as a metrics in order to evaluate the performance of the different protocols:

- **Average end-to-end delay**: The delay of a packet is the time it takes the packet to achieve the destination after it leaves the source. The average packet delay for a network is obtained by averaging over all packets and all source destination pairs. The average end-to-end delay $T_{Avg}$ is calculated as showing in equation (1):

$$T_{AVG} = \frac{\sum_{i=1}^{Nr}\left(H_r^i - H_t^i\right)}{Nr} \qquad (1)$$

$H_t^i$ emission instant of package $i$, $H_r^i$ reception instant of package $i$, $N_r$ the total number of packets received

- **Throughput**: The ratio of successfully transmitted data per second (2).

$$T = \tfrac{L-C}{L} Rf(\gamma) \qquad (2)$$

Where $\tfrac{L-C}{L} Rf(\gamma)$ is the payload transmission rate, *(R)* b/s Binary transmission rate, *(L)* Packet size, and $f(\gamma)$ is the packet success rate defined as the probability of receiving a packet correctly. This probability is a function of the signal-to-noise ratio $(\gamma)$.

- **Packet Delivery Ratio**: The ratio of the data packets successfully delivered to the destination.

$$\frac{\sum\left(\text{Number of Received Data Packets}\right)}{\sum\left(\text{Number of Sent Data Packets}\right)}$$

## 5. RESULTS DISCUSSION

In this section we present our simulation results and the performance analysis. The analysis based on comparing the different metrics of the mobility models that we described previously in section 3.

### 5.1. Variable Bit Rate (VBR)

As showing in Figure 1, with the three mobility models, the delay increased when increasing density of nodes. So with Random Direction, OLSR take less time to deliver the packets compared to the two other models (Random Way Point, Mobgen Steady State).

On the other hand, the most popular mobility model used in the literature is Random Way Point [29]. From Figure 1, an important point is Mobgen Steady State give best performance than Random Way point in terms of delay. Because the Mobgen Steady State is more realistic than the Random Direction model the optimal delay (refer Figure 1) is achieved with Mobgen Steady State. Hence we promotes the use of the Mobgen Steady State model in the applications that sensitive to delay.

17

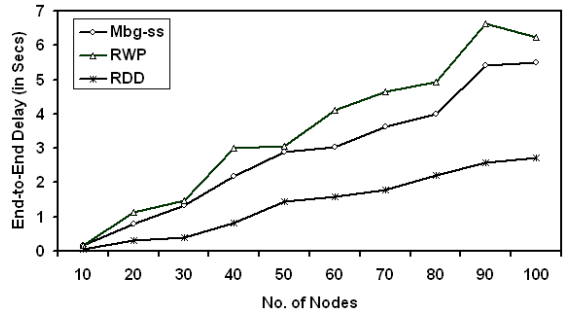

Figure 1. End-to-End Delay vs. No. Of Nodes with VBR.

Based on Figure 2, with Random Way Point, OLSR shows higher throughput than both Random Direction and Mobgen Steady State when weak density of nodes is used. So Random Way point produces a high throughput more than, respectively, both Mobgen Steady State and Random Direction in the first part. However, in the second part Random Way Point and Random Direction outperforms than Mobgen Steady State and they gives the same behaviour.

On the other side if we consider the applications that require a certain level of throughput (refer Figure 2) we suggest using Random Way Point in weak densities of nodes. For the same applications, when using heavy densities of knots we can choice between both Random Way Point and Random Direction.

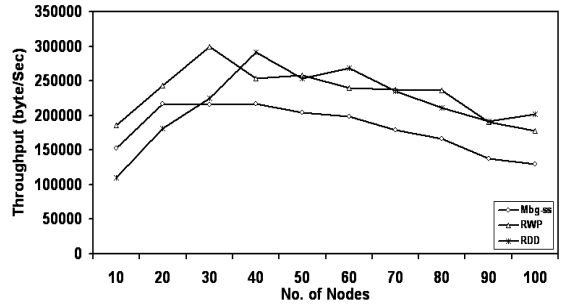

Figure 2. Throughput vs. No. Of Nodes with VBR.

As showing in the Figure 3 a higher packet delivery ratio for small density, is achieved when using OLSR with Random Way Point mobility model. Finally, based on behaviour of variability of VBR (MPEG-4), the Packet Delivery Ratio still insufficient over all density of nodes. This for all the three mobility models especially with heavy density is used. Hence, OLSR protocol can be used on applications that tolerate a small amount of packet loss.

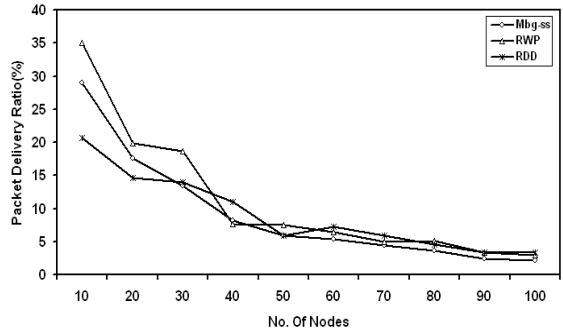

Figure 3. Packet Delivery Ratio vs. No. Of Nodes with VBR.

## 5.2. Constant Bit Rate (CBR)

Figure 4 shows that the performance of the OLSR routing protocol in terms of end-to-end delay increase by increasing density of knots. With Random Direction mobility model, OLSR take less time to deliver the packets compared to Random Way Point and Mobgen Steady State mobility models. Because the same reasons, Mobgen Steady State is an improvement of Random Way point and this latest is more realistic than Random Direction, we can tell that the optimal delay is achieved by Mobgen Steady State.

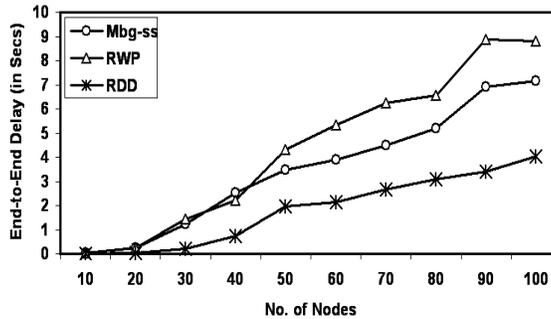

Figure 4. End-to-End Delay vs. No. Of Nodes with CBR.

It is expected that if the node density is increased the throughput of the network shall increase. However, in the first (refer Figure 5) part throughput of all mobility models increase when increasing number of nodes, because the number of load is small and the traffic is not heavy. Beyond this part, the throughput with Random Way Point decreases and it's still stable and consistent for the three mobility models.

If we consider the applications that require a certain level of throughput the results suggest using OLSR in case of CBR traffic with Random Way Point.

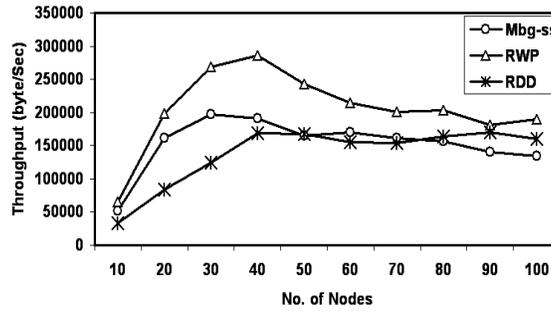

Figure 5. Throughput vs. No. Of Nodes with CBR.

Based on Figure 6, with Random Way Point mobility model, OLSR performed better in delivering packet data to the destination than the two others. As showing (refer Figure 6) the best performance in terms of Packet Delivery Ratio is got in small density of knots. However, when density becomes heavy the Packet Delivery Ratio still insufficient with all mobility models considered.

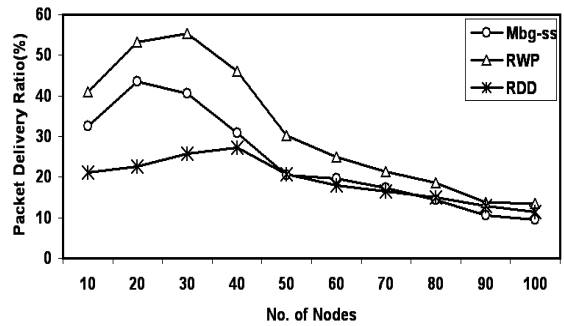

Figure 6. Packet Delivery Ratio vs. No. Of Nodes with CBR.

### 5.3. VBR and CBR

The first remark when changing traffic from CBR to VBR (MPEG-4) on the performance of OLSR routing protocol in terms of end-to-end delay is that OLSR keeps the same behaviour (delay increase when increasing density of knots) with the three mobility models considered. On the other hand the most popular mobility model used in the literature is Random Way Point [29]. But as showing on (figures 1 and 4) Mobgen Steady State performed better than Random Way Point.

Based on figures 2 and 5 the OLSR protocol is not influenced when changing traffic from CBR to VBR (MPEG-4) in weak density of knots. Over all densities used, if we consider that Random Direction is not performed in term of Packet Delivery Ratio (refer Figures 3 and 6), the Random Way Point still the best mobility model to use especially for the applications that need a certain level of throughput.

Despite the acceptable behaviour showed by OLSR in case of small densities (refer figure 6), the Packet Delivery Ratio still insufficient over all density (refer figures 3 and 6) of nodes without considering the traffic type (CBR and VBR (MPEG-4)). This for all the three mobility models. Hence, we promote to use OLSR protocol associated with Random Way Point on applications that tolerate a small amount of packet loss.

## 6. CONCLUSION AND FUTURE WORK

In this paper we have conducted a behaviour study of OLSR routing protocol using two traffic types multimedia (VBR) and CBR, by over various mobility models as Random Way Point, Random Direction and Mobgen Steady State.

With OLSR model in association with CBR traffic, in the first one, the optimal delay is achieved respectively by Random Way Point in small density and Mobgen Steady State in heavy density. In the second one, the optimal throughput is achieved by Random Way Point.

In the association of OLSR model with traffic VBR (MPEG-4), the optimal delay is got by means of Mobgen Steady State. However, the optimal throughput is achieved by Random Way Point.

The less packet delivery under VBR traffic is due to the proactive nature of OLSR routing protocol, on the variability of the VBR traffic and to the high mobility caused by pause time value (0 second). However, in the second case when using CBR traffic it's due to the proactive nature of OLSR routing protocol.

There is no way to tell that a particular mobility model is well versed in all types of scenarios. The selection of a mobility model and routing protocol is based on the final application of Ad Hoc Network. With this study, we hope to help the future studies in their choice of parameters and models. This, in order to design the realistic scenarios which depict real world applications more accurately and more of QoS.

Other most important point in this paper is the behaviour of OLSR, with the three mobility described previously, depend on the traffic used (CBR or VBR). This behaviour is influenced precisely in case of low densities of nodes.

One of the most interesting parameters to consider when supporting real time communication is the delay jitter. In the future, further study also needs to be done with delay jitter metric.

On the other hand, in the future, further study should be devoted to optimize the Packet Delivery Ratio when using traffic VBR.